\documentclass[11pt,twoside]{article}

\usepackage{asp2006}
\usepackage{epsfig}
\usepackage{psfig}
\usepackage{lscape}
\usepackage{wrapfig}

\begin{document}
\title{Dark Matter in Gas-Rich Dwarf Galaxies}   
\author{U. Klein$^1$, G. Gentile$^2$, P. Salucci$^3$, G. J, G. J\'ozsa$^1$, 
and F. Kenn$^1$}   
\affil{1: AIfA, Bonn, 2: UNM, Albuquerque, 3: SISSA, Trieste}    

\begin{abstract} 
We investigate the distribution of dark matter (DM) in gas-rich, low-mass 
galaxies, confronting them with numerical cosmological simulations with 
cold dark matter ($\Lambda$CDM). We show that the derived rotation curves 
comply best with cored DM density profiles, whereas the signatures
of the central cusps invariably predicted by $\Lambda$CDM simulations 
are not seen. 
\end{abstract}



\section{Introduction}

Rotation curves of disk galaxies constitute the most powerful tools to 
study dark matter, its content relative to baryons, and its distribution 
on the scale of galaxies. In particular, dwarf galaxies are prime 
candidates for DM studies as their kinematics is generally dominated 
by dark matter down to small galactocentric radii (Persic et al. 1996), 
Thus, they provide an important test for cosmological models, because 
numerical cosmological simulations with CDM predict a common 
characteristics, a central steep rise (cusp) in the DM density 
profile (e.g. Navarro, Frenk and White 1996). Following a thorough 
investigation of a sample of low-luminosity galaxies (Gentile et al. 
2004), which clearly documents the problems with central cusps as 
described above, we have embarked on a study of dwarf galaxies that 
are gas-rich, isolated and DM dominated.

\section{The targets}

DDO\,47 is a dwarf irregular galaxy at 4~Mpc distance. The galaxy 
has a fairly regular structure. The velocity field of the gas looks 
regular, rendering this galaxy a convenient object for kinematic 
studies. NGC\,3741 is a dwarf irregular galaxy at 3.5~Mpc distance. 
The HI observations reveal the largest disk in terms of optical 
scale length, viz. 42 $\times ~ R_{opt}$. This allows us to probe 
the galactic potential out to exceptionally large radii. DDO~154, 
an extremely gas-rich dwarf galaxy at 3.2~Mpc distance, is known 
to have a large gaseous disk and a very asymmetric rotation curve, 
which appears to decline in the outer parts (e.g. Carignan \& Purton 
1998). 

The HI data cubes have been used to thoroughly investigate the 
kinematics of the galaxies (Gentile et al 2005, 2007). Apart from 
deriving their rotation curves (s.b.), this also involved a harmonic 
analysis, such as to uncover non-circular motions and asymmetries 
in the kinematics of the gaseous disks. In DDO\,47, non-circular 
motions of below 3~km~s$^{-1}$ were detected, likely connected with 
the spiral structure. NGC\,3741 exhibits non-circular motions of up 
to 10~km~s$^{-1}$, most likely due to a bar in the inner region and 
to gas inflow in the outer one. DDO\,154 exhibits clear signs of 
non-circular motions. We have analyzed the data cube with our 
new tilted-ring fitting code (TiRiFiC, J\'ozsa et al. 2007; see 
also J\'ozsa et al., these proceedings). Our preliminary 3-D 
reconstruction of the gaseous disk using TiRiFiC manifests the 
complex morphology and kinematics, its systemic velocity and 
centroid varying as a function of radius. The latter could be 
the signature of recent gas infall, giving rise to an unrelaxed 
baryonic disk; this is corroborated by the non-circular motions, 
which can be interpreted in terms of gas infall.

\section{Rotation curves}

\begin{wrapfigure}{r}{5cm}
\includegraphics[width=5cm]{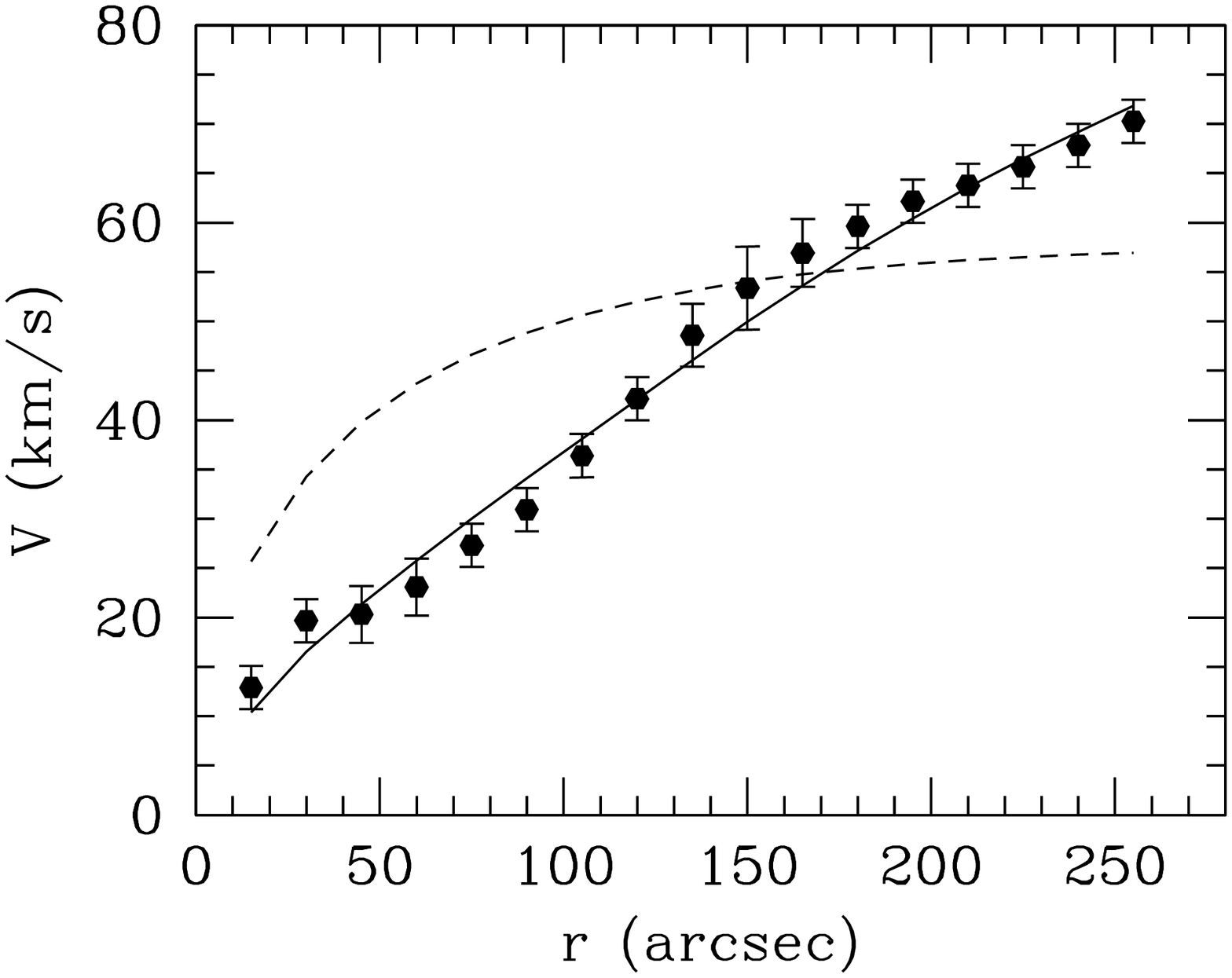}
\end{wrapfigure}
We have used the rotation curves of DDO\,47 and NGC\,3741 to perform a 
mass decompsition. The resulting circular velocities of the DM halo 
were compared with models. The Figure shows mass models for DDO 47, 
with the solid line representing the best-fitting model with a Burkert 
halo, while the dotted line represents the best-fitting NFW model.
It is obvious that the rotation curve of DDO\,47 is in conflict with 
an NFW law, while a Burkert law yields a rather good fit. The kinematic 
signatures found along the minor axis are too small to wash out any 
cusp-shaped density profile, thus mimicking a cored DM halo as proposed 
by Hayashi et al. (2004). The rotation curve of NGC\,3741, too, fits the 
Burkert halo very well, while $\Lambda$CDM halo profiles produce worse 
fits. MoND also delivers acceptable fits. The derivation of the rotation 
curve of DDO\,154 has to await the final analysis with TiRiFiC.

\section{Conclusions}

In summary, it is obvious that the rotation curves of DDO\,47 and 
NGC\,3741 are easily reconciled with cored density distributions of their 
DM halos, while $\Lambda$CDM is still facing a serious problem here.

\acknowledgements 
This work has financial support by the Deutsche For\-schungsgemeinschaft
(grant KL~533/8-2).


\end{document}